\documentclass[aps,prl,twocolumn,showpacs,superscriptaddress,dblfloatfix,preprintnumbers]{revtex4-1}
\usepackage{graphicx}   
\usepackage{dcolumn}   
\usepackage{bm}           
\usepackage{amssymb}  
\usepackage{amsmath}
\usepackage{dsfont}
\usepackage{setspace}
\usepackage{amsmath, amssymb, setspace}
\usepackage{array}
\usepackage{booktabs}
\usepackage{mathrsfs}
\usepackage{indentfirst}
\usepackage{slashed}
\usepackage{float}
\usepackage{lmodern}
\usepackage{hyperref}
\usepackage{xcolor}
\usepackage{multirow}
\usepackage{epstopdf}
\usepackage{ulem}
\usepackage{tabularx}
\usepackage[justification=raggedright]{caption}
\usepackage[flushleft]{threeparttable}
\usepackage{hyperref}
\usepackage{soul}
\usepackage{braket}

\begin{document}


\preprint{IPMU16-0192}

\title{Primordial black holes from supersymmetry in the early universe}

\author{Eric Cotner}
\affiliation{Department of Physics and Astronomy, University of California, Los Angeles\\
Los Angeles, CA 90095-1547, USA}

\author{Alexander Kusenko}
\affiliation{Department of Physics and Astronomy, University of California, Los Angeles\\
Los Angeles, CA 90095-1547, USA}
\affiliation{Kavli Institute for the Physics and Mathematics of the Universe (WPI), UTIAS\\
The University of Tokyo, Kashiwa, Chiba 277-8583, Japan}

\date{\today}

\begin{abstract}
Supersymmetric extensions of the standard model generically predict that in the early universe a scalar condensate can form and fragment into Q-balls before decaying.  If the Q-balls dominate the energy density for some period of time, the relatively large fluctuations in their number density can lead to formation of primordial black holes (PBH). Other scalar fields, unrelated to supersymmetry, can play a similar role.  For a general charged scalar field, this robust mechanism can generate black holes over the entire mass range allowed by observational constraints, with a sufficient abundance to account for all dark matter in some parameter ranges.  In the case of supersymmetry the mass range is limited from above by $10^{23}$g.  We also comment on the role that topological defects can play for PBH formation in a similar fashion.
\end{abstract}


\maketitle

It is a long-standing question whether black holes could form in the early universe~\cite{Zeldovich:1967,Hawking:1971ei,Carr:1974nx,GarciaBellido:1996qt,Khlopov:2008qy,Frampton:2010sw,Kawasaki:2012kn,Kawasaki:2016pql,Carr:2016drx,Inomata:2016rbd,Inomata:2017okj,Georg:2017mqk}.  Primordial black holes (PBH) could account for all or part of dark matter~\cite{Zeldovich:1967,Hawking:1971ei,Carr:1974nx,GarciaBellido:1996qt,Khlopov:2008qy,Frampton:2010sw,Kawasaki:2016pql,Carr:2016drx,Inomata:2016rbd,Inomata:2017okj,Georg:2017mqk}, they could be responsible for some of the gravitational wave signals observed by LIGO~\cite{Nakamura:1997sm,Clesse:2015wea,Bird:2016dcv}, and they could provide seeds for supermassive black holes~\cite{Kawasaki:2012kn}. 
A number of scenarios for black hole formation have been considered~\cite{Khlopov:2008qy}, and many of them rely on a spectrum of primordial density perturbations that has some extra power on certain length scales, which can be accomplished by means of tuning an inflaton potential.  

In this \textit{Letter} we will present a more generic scenario for PBH formation in the early universe, which does not rely on any particular spectrum of density perturbations from inflation.  Scalar fields with slowly growing potentials form a coherent condensate at the end of inflation~\cite{Bunch:1978yq,Linde:1982uu,Affleck:1984fy,Starobinsky:1994bd}.  In general, the condensate is not stable, and it breaks up in lumps, which evolve into Q-balls~\cite{Kusenko:1997si}. The gas of Q-balls contains a relatively low number of lumps per horizon, and the mass contained in these lumps fluctuates significantly from place to place.  This creates relatively large fluctuations of mass density in Q-balls across both subhorizon and superhorizon distances. Since the energy density of a gas of Q-balls redshifts as mass, it can come to dominate the energy density temporarily, until the Q-balls decay, returning the universe to a radiation dominated era. The growth of structure during the Q-ball dominated phase can lead to copious production of primordial black holes.  

Formation of Q-balls requires nothing more than some scalar field with a relatively flat potential at the end of inflation.  For example, supersymmetric theories predict the existence of scalar fields with flat potentials.  PBH formation in supersymmetric theories is, therefore, likely, even if the scale of supersymmetry breaking exceeds the reach of existing colliders. 

A similar process can occur with topological defects, which can also lead to relatively large inhomogeneities.  The discussion of topological defects is complicated by their non-trivial evolution.  We will focus primarily on Q-balls, and will briefly comment on topological defects below. 

\textit{Formation of Q-balls} occurs by fragmentation of a scalar condensate after inflation~\cite{Kusenko:1997si}.  While supersymmetry is a well-motivated theory for scalar fields carrying global charges and having flat potentials\cite{Affleck:1984fy,Dine:2003ax}, our discussion can be easily generalized to an arbitrary scalar field with a global U(1) symmetry in the potential. Supersymmetric potentials generically contain flat directions that are lifted only by supersymmetry breaking terms.  Some of the scalar fields that parameterize the flat directions carry a conserved U(1) quantum number, such as the baryon or lepton number.  During inflation, these field develop a large vacuum expectation value (VEV)~\cite{Bunch:1978yq,Linde:1982uu,Affleck:1984fy,Starobinsky:1994bd}, leading to a large, nonzero global charge density. When inflation is over, the scalar condensate $ \phi(t) = \phi_0(t) \exp \{i\theta (t)\}$ relaxes to the minimum of the potential by a coherent classical motion with $\dot\theta\neq 0$ due to the initial conditions and possible CP violation at a high scale. 

The initially homogeneous condensate is unstable with respect to fragmentation into non-topological solitons, Q-balls~\cite{Coleman:1985ki}. Q-balls exist in the spectrum of every supersymmetric generalization of the Standard Model~\cite{Kusenko:1997zq,Kusenko:1997ad}, and they can be stable or long-lived along a flat direction~\cite{Kusenko:1997si,Dvali:1997qv}. In the case of a relatively large charge density (which is necessary for Affleck-Dine baryogenesis \cite{Affleck:1984fy,Dine:2003ax}), the stability of Q-balls can be analyzed analytically \cite{Kusenko:1997si,Enqvist:1997si,Enqvist:1998en}; these results agree well with numerical simulations \cite{Kasuya:1999wu}. One finds that the almost homogeneous condensate develops an instability with wavenumbers in the range $0<k<k_\text{max}$, where $k_\text{max}=\sqrt{\omega^2 - V''(\phi_0)} $, and $\omega=\dot\theta$.  
The fastest growing modes of instability have a wavelength $\sim 10^{-2\pm 1}$ of the horizon size at the time of fragmentation, and they create isolated lumps of condensate which evolve into Q-balls.  Numerical simulations~\cite{Kasuya:1999wu,Kasuya:2000wx} indicate that most of the condensate ends up in lumps. However, since the mass of Q-balls is a non-linear function of the Q-ball size, Q-ball formation, in general, leads to a non-uniform distribution of energy density in the matter component represented by the scalar condensate.  Q-balls can also form when the charge density is small or zero, in which case both positively and negatively charged Q-balls are produced~\citep{Kasuya:1999wu}; here we do not consider this possibility.

If the potential is flat, a Q-ball with global charge $Q$ has mass $M \propto |Q|^{3/4}$~\cite{Dvali:1997qv,Kusenko:1997si}.  The density fluctuations arise from the non-linear relation between $M$ and $Q$.   In general, $M\sim |Q|^\alpha$, where $\alpha$ depends on the potential.
If $\alpha = 1$, some density fluctuations may arise from global charge redistribution during fragmentation. We do not consider this possibility here, and we limit our discussion to $\alpha=3/4$.

\textit{Q-ball number distribution.} Let us consider $N$ identical Q-balls in some  volume $V$ at the time of fragmentation $t_f$. We assume that the  probability to find $N$ Q-balls in a given volume $V$ follows a Poisson distribution: 
\begin{gather}
p(N,V) = e^{-(N_f V/V_f)} \frac{(N_f V/V_f)^N}{N!}, \quad N \in \mathds{Z}^+,
\end{gather}
where $N_f$ is the average number of Q-balls per horizon (within volume $V_f=\frac{4\pi}{3} t_f^3$, the horizon volume at fragmentation). The mass of a volume containing $N$ Q-balls is given by $M(N) = N M_\text{Q-ball} = \Lambda N Q^\alpha$. If this mass is large enough, it becomes the mass of the black hole resulting from the collapse. 
The charge within the volume $V$ at fragmentation is assumed to be distributed equally among the $N$ Q-balls,  so that $N Q = Q_f V/V_f$ (with $Q_f$ the total charge on the horizon at $t_f$), which gives us the Q-ball cluster distribution function $F_Q$
\begin{gather} \label{eq:QballDistribution}
F_Q(M,N,V) = \delta\left(M - M_f \left(\frac{N}{N_f}\right)^{1-\alpha} \left(\frac{V}{V_f}\right)^\alpha\right) p(N,V),
\end{gather}
where $M_f = \Lambda Q_f^\alpha N_f^{1-\alpha}$ is the average Q-ball horizon mass at $t_f$. $F_Q$ represents the probability density to find a mass $M$ composed of $N$ Q-balls within a volume $V$.

The average background energy density (over the largest scales) in Q-balls at $t_f$ is then given by
\begin{gather}
\braket{\rho_Q(t_f)} = \lim\limits_{V\rightarrow\infty} \frac{\braket{M}}{V} = \frac{M_f}{V_f},
\end{gather}
with the average performed over $M$ and $N$.

\begin{figure}
\centering
\includegraphics[width=0.9\linewidth]{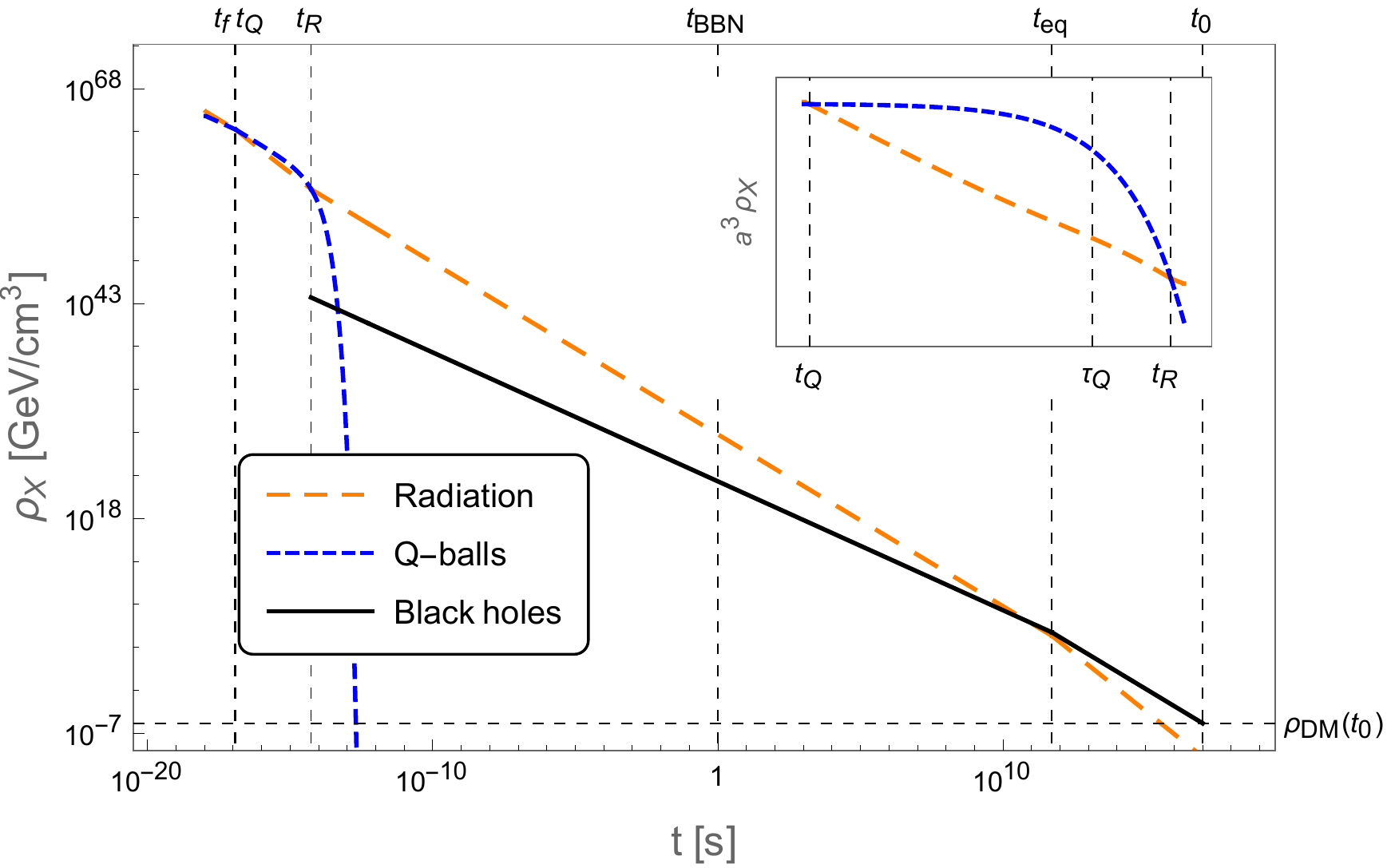}
\caption{Cosmological timeline corresponding to the production of PBH. Orange dashed line denotes radiation density $\rho_R$, blue dashed line is Q-ball energy density $\braket{\rho_Q}$, and black solid line is black hole density $\braket{\rho_\text{BH}}$. Inset in the upper right is a zoomed-in view of the early matter dominated era ($t_Q < t < t_R$), with density scaled by $a^3$ (so that non-decaying matter would be represented by a straight horizontal line). Horizontal dashed line indicates the observed present-day dark matter density. Here the parameters correspond to the solid line in Fig.~\ref{fig:CombinedConstraintPlots}.}
\label{fig:CosmologicalTimelineGrid002}
\end{figure}

Q-balls are stable with respect to decay into scalar particles, but they can decay into fermions lighter than $\omega$~\cite{Cohen:1986ct,Kawasaki13,Enqvist:1998xd}.  Q-balls can also decay if the U(1) symmetry is broken by some higher-dimension  operators~\cite{Kusenko:2005du,Kawasaki:2005xc,Kasuya:2014ofa,Cotner:2016dhw}.  We parameterize this decay by the total decay width $\Gamma_Q=1/\tau_Q$, which includes all decay channels. The energy density in the form of Q-balls scales with expansion of the universe as decaying matter~\cite{Scherrer:1984fd}, and the decays also contribute to radiation density.  We take this into account in a consistent manner using the analysis of Scherrer and Turner \cite{Scherrer:1984fd}. The energy density of Q-balls then evolves as $\braket{\rho_Q(t)} = \braket{\rho_Q(t_f)} (a_f/a)^3 e^{(t_f - t)/\tau_Q}$, and we assume that at some time $t_Q$ the Q-balls come to dominate, as shown in Fig.~\ref{fig:CosmologicalTimelineGrid002}.

\textit{Density perturbations} can grow starting at time $t_Q$ when the universe becomes matter dominated (i.e., $\rho_R(t_Q) = \braket{\rho_Q(t_Q)}$) until the time when the Q-balls decay, returning the universe to a radiation dominated phase. The structures can grow on all length scales above some minimum size $V_\text{min}$, which we take to be the volume containing an average number of Q-balls $N_\text{min} \sim 10$. This acts as a cutoff to the low-mass part of the PBH spectrum, but does not influence any larger scales, and we have checked the results are not sensitive to this choice as long as $N_\text{min} \ll N_f$.

The density contrast in Q-balls at fragmentation $\delta(t_f)$ of a specific mass scale $M$ composed of $N$ Q-balls within a volume $V$ at $t_f$ is given by
\begin{gather}
\delta(t_f)  = \frac{\delta\rho}{\braket{\rho_Q}} = \frac{M/V}{\braket{\rho_Q}} - 1 = \left(\frac{N/N_f}{M/M_f}\right)^\frac{1-\alpha}{\alpha} - 1
\end{gather}
where in the last line we have used the argument of the delta function in Equation \ref{eq:QballDistribution} to eliminate $V$. Note that when $\alpha = 1$, the density perturbations vanish identically.

The density perturbations are frozen during the radiation dominated era, but they grow linearly in the scale factor during the Q-ball dominated epoch, $\delta(t) = \delta(t_f) (a/a_Q) = \delta(t_f) (t/t_Q)^{2/3}$. The  structure growth generally goes nonlinear and decouples from the expansion around $\delta > \delta_c \sim 1.7$, at which point the overdense regions collapse and become gravitationally bound. However, some structures with $\delta < \delta_c$ can still collapse, and not all structures with $\delta > \delta_c$ are guaranteed to collapse into black holes. Due to nonsphericity of the gravitationally-bound structures, only a fraction $\beta = \gamma \delta^{13/2}(t_R) (M/M_Q)^{13/3}$ (where $\gamma \approx 2 \times 10^{-2}$ and $M_Q=M_f(t_Q/t_f)^{3/2}$ is the horizon mass at the beginning of the Q-ball dominated era) will actually collapse spherically to form black holes~\cite{Polnarev:1986bi,Carr:2016drx,Georg:2016yxa} by the end of the Q-ball dominated era ($t_R$). Structures with $\delta \ge \delta_c$ do not continue to grow (as perturbations are gravitationally bound at this point and cease developing), so that $\beta = \gamma \delta_c^{13/2} (M/M_Q)^{13/3}$ for $\delta(t_R) > \delta_c$. Despite these refinements, the outcome does not appear to depend sensitively on the value of $\delta_c$.

Additional care must be taken to extend this to scales which enter the horizon during the Q-ball dominated era, and thus are not subject to the same amount of growth as the subhorizon modes. This can be done by calculating the time $t_h$ at which a comoving superhorizon volume $V$ at $t_f$ re-enters the horizon: $(a(t_h)/a_f)^3 V = V_h = \frac{4\pi}{3} t_h^3$. Then, the amplification of these superhorizon modes at the end of the Q-ball dominated era are is given by $\delta(t_f)(a_R/a(t_h))$ rather than $\delta(t_f)(a_R/a(t_Q))$. Also, because this expression is only valid for small $\delta$, we will cap its value at $\beta_\text{max} = 1$ to avoid collapse probabilities over unity. Structure growth ends once the radiation density comes to dominate again at time $t_R$, which is defined by $\rho_R(t_R) = \braket{\rho_Q(t_R)}$.

\textit{PBH production} in this mechanism can be analyzed by first calculating the energy density of Q-balls at $t_f$ that \textit{will eventually} form black holes by $t_R$ by weighting the Q-ball energy density $M/V$ by the collapse fraction/probability $\beta$ evaluated at $t_R$, and then redshifting this value appropriately. In addition, one must sum the contributions of all length scales $V$ through a coarse-graining procedure. This is accomplished for some arbitrary function $g(V)$ via the procedure
\begin{gather}
\sum_{\{V\}} g(V) = g(V_1) + g(V_1/\chi) + \cdots = \sum_{i=1}^{I_\text{max}} g(V_1\chi^{1-i}) \\
\approx \int_{1}^{I_\text{max}} di\, g(V_1\chi^{1-i}) = \frac{1}{\ln\chi} \int_{V_\text{min}}^{V_1} \frac{dV}{V} g(V),
\end{gather}
where we have used the Euler-Maclaurin summation approximation, $\chi \sim O(1-10)$ is a parameter of the coarse-graining, and $V_\text{min}$ is the smallest volume scale under consideration, set by $N_\text{min}$. We will henceforth take $\chi = e$ for simplicity.

At $t_R$, the black hole density is given by 
\begin{align} \label{eq:BHdensity}
\braket{\rho_\text{BH}(t_R)} &= \left(\frac{a_f}{a_R}\right)^3 \sum_{N=1}^{\infty} \int_{V_\text{min}}^{V_R} \frac{dV}{V} \int_{0}^{\infty} dM\, \left(\beta \frac{M}{V}\right) F_Q
\end{align}
where $V_R = \frac{4\pi}{3} t_R^3 (t_Q/t_R)^2 (t_f/t_Q)^{3/2}$ is the size of the comoving horizon at $t_R$, evaluated at $t_f$. The black hole energy density then redshifts like nonrelativistic matter, $\braket{\rho_\text{BH}(t)} = \braket{\rho_\text{BH}(t_R)} (a_R/a(t))^3$.

\begin{figure}
\centering
\includegraphics[width=0.9\linewidth]{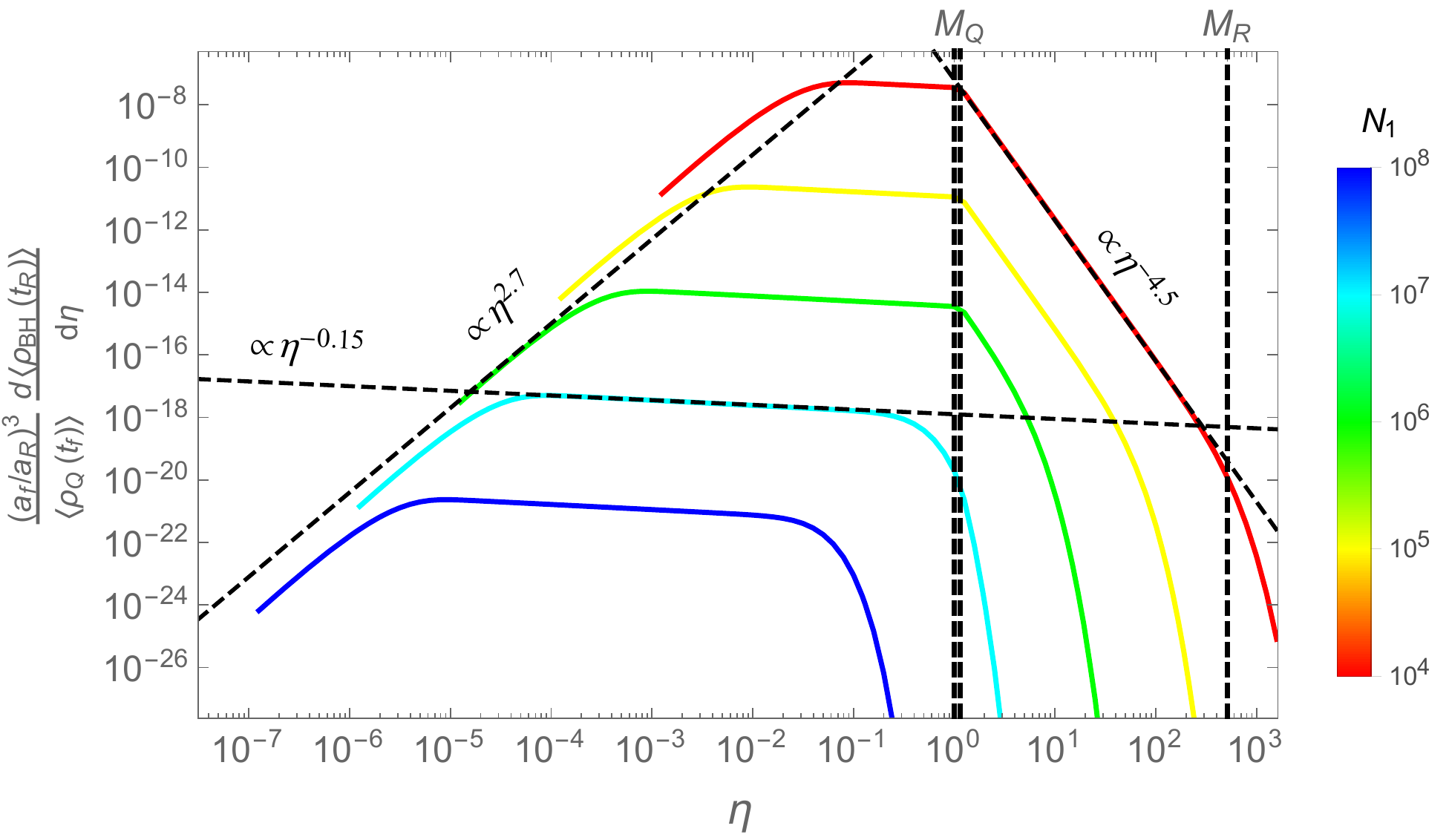}
\caption{Differential fraction of Q-ball energy density transferred to BH density as a function of $\eta=M/M_f$. This spectrum corresponds to the parameters for the solid black line in Figure \ref{fig:CombinedConstraintPlots}, which accounts for 100\% of the dark matter.} 
\label{fig:MassSpectrum}
\end{figure}

\textit{The mass function} $d\braket{\rho_\text{BH}}/dM$ can be evaluated using the integrand of Eq.~\ref{eq:BHdensity} ($\braket{\rho_\text{BH}} = \int dM\, d\braket{\rho_\text{BH}}/dM$). Using this ``differential density" mass function, shown in Fig.~\ref{fig:MassSpectrum}, one can glean some information regarding how the parameters of the theory affect the distribution of black hole mass. We find that the spectrum depends only on the dimensionless parameters $N_f$, $\eta = M/M_f$, $r_f = t_Q/t_f$, and $r=t_R/t_Q$ (given this set, plus a value for $t_f$, fixes the remaining parameters $\tau_Q$, $M_f$ due to consistency conditions on the boundary of the Q-ball dominated era). First, it's obvious from the normalization of each curve that the lower the number of Q-balls per horizon, the more black holes that are created. This is expected, as the Poisson statistics suppress the density fluctuations for large Q-ball number. Second, there is a hard lower cutoff in the PBH mass, which occurs at $\eta = N_\text{min}/N_f$. Above that, the BH number sharply increases with a power law $\propto \eta^{2.7}$; the extent of this region depends on the magnitude of $r$, with larger values leading to a larger range. Above that, the spectrum becomes approximately flat ($\propto \eta^{-0.15}$), meaning that the number of black holes in each decade of mass are comparable. Of course, the upper end of this range dominates the energy density of the distribution. Then, at around $M = M_Q$, there is a sharp transition and the slope becomes strongly negative ($\propto \eta^{-4.5}$) due to the reduced growth the superhorizon modes are subject to. Then, there is an upper exponential cutoff at $\eta \sim 10^8/N_f$ due once again to the Poisson statistics (the cutoff appears to take precedence over previously mentioned transitions). For the parameters given in Figure \ref{fig:MassSpectrum}, the spectrum is highly developed,  in the sense that it has been subject to a lengthy matter dominated era of growth ($r \gg 1$).

\textit{Experimental constraints} can be considered after we evolve the black hole distribution to the present day. In simple terms, we just have to take Equation \ref{eq:BHdensity} and multiply it by $(a(t_0)/a(t_R))^{-3}$. Na\"ively, one would use the equation $a_1/a_2 = (t_1/t_2)^n$ (with $n=1/2$ or $2/3$) keeping in mind that the Universe transitions back to a matter-dominated era around $z_\text{eq} \approx 3360$. However, an extended Q-ball dominated era ($r \gg 1$) alters the timescale of cosmological thermal history because the radiation temperature is altered by the Q-ball decays and the form of the scale factor during this era. In this case, one must use $a_1/a_2 = g_{*S}^{1/3}(T_2) T_2/g_{*S}^{1/3}(T_1) T_1$ and evolve from $T_R$ (defined by $\rho_R(t_R) = (\pi^2/30) g_{*}(T_R) T_R^4$) to $T_0 = 2.7 \text{ K} = 2.3 \text{ meV}$. This has the advantage of accurately accounting for any deviation from cosmological history. In addition, we enforce an additional constraint $T_R > T_\text{BBN} \sim \text{MeV}$, so that the entropy injection from Q-ball decays does not interfere with nucleosynthesis.

\begin{figure}
\centering
\includegraphics[width=0.9\linewidth]{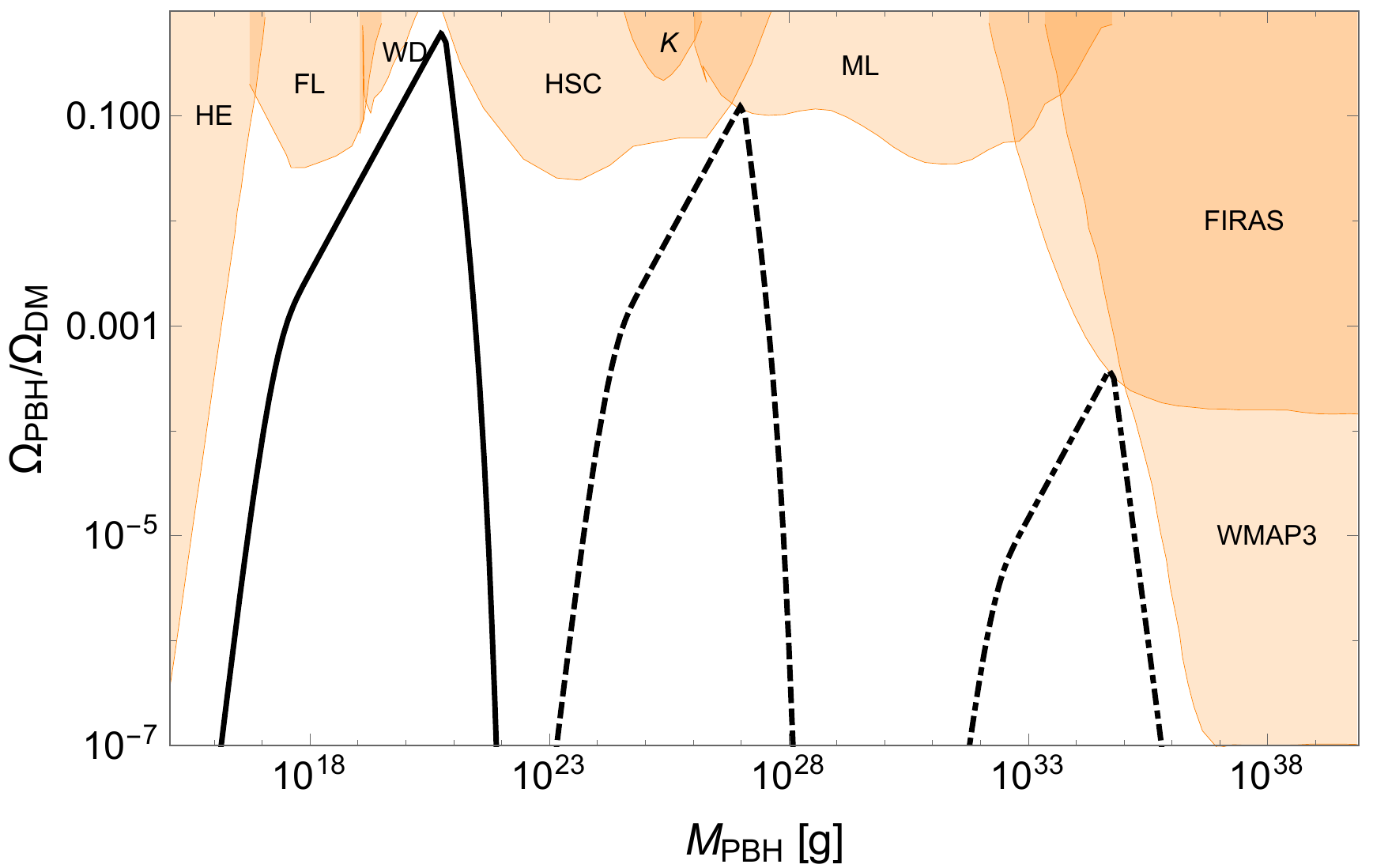}
\caption{Comparison of the observational constraints on $f \equiv \Omega_\text{PBH}/\Omega_\text{DM}$ (orange, shaded), with the expected value of $\tilde{f}_\text{BH}(M) \equiv \rho_\text{DM}^{-1} d\braket{\rho_\text{BH}}/d(\ln M)$ for some hand-picked parameters (black). Parameters for the three curves are $t_f = 1.12\times 10^{-17}\text{ s}$, $r_f=1.1$, $r=4.47\times 10^2$, $N_f=10^6$, $f=1$ (solid line), $t_f=2.0\times 10^{-11} \text{ s}$, $r_f = 1.1$, $r=1.58\times 10^3$, $N_f = 10^6$, $f=0.2$ (dashed line), and $t_f=1.0\times 10^{-3}\text{ s}$, $r_f=1.1$, $r=4.47\times 10^2$, $N_f=10^5$, $f=0.001$ (dot-dashed line).}
\label{fig:CombinedConstraintPlots}
\end{figure}

We now apply observational constraints to our model. There are a variety of constraints \cite{Carr:2009jm,Kuhnel:2017pwq,Carr:2016drx,Niikura:2017zjd,Inomata:2017okj,Inoue:2017}, coming from a number of sources, including gamma rays from Hawking radiation, femto-, micro-, and milli-lensing, white dwarf capture, pulsar timing arrays, and accretion effects on the CMB. These constraints can be observed in Figure \ref{fig:CombinedConstraintPlots} in orange. It is important to note that the constraints as plotted only apply for a monochromatic mass distribution. The mass spectrum of this model clearly extends over several decades, and so we must translate these constraints into something applicable to our model. We adopt the procedure outlined in \cite{Carr:2016drx}, which amounts to comparing the expected dark matter fraction to the constraints on an interval-by-interval basis. As a guide, we also plot $\tilde{f}_\text{BH}(M) \equiv \rho_\text{DM}^{-1} d\braket{\rho_\text{BH}}/d(\ln M)$, which gives an approximate idea of how the expected BH density in each logarithmic interval of mass compares with the constraints. We have verified for the given parameters that the constraints have not been violated. We note that the solid curve corresponds to a distribution that makes up 100\% of the dark matter with peak black hole mass $10^{20} \text{ g}$. For the case of supersymmetric Q-balls with the SUSY-breaking scale $\Lambda_\text{SUSY} > 10\text{ TeV}$, the fragmentation time cannot be much longer than the Hubble time $H^{-1} \sim M_p/g_*^{1/2} \Lambda_\text{SUSY}^2 \lesssim 8\times 10^{-15} \text{ s}$, which corresponds to peak PBH masses of about $10^{23} \text{ g}$. The case illustrated in Figure \ref{fig:CombinedConstraintPlots} satisfies this bound, thus primordial black holes from supersymmetric Q-balls can account for 100\% of the dark matter.

The dot-dashed curve corresponds to a distribution that only makes up 0.1\% of the dark matter, but has a peak BH mass of $30 \text{ M}_\odot$. This is suggestive, as even if the dark matter isn't entirely PBHs, they might still be responsible for some of the black hole merger events detected by LIGO~\cite{Bird:2016dcv}.

\textit{Topological defect} formation can also lead to the production of PBHs if the topological defects come to dominate the energy density. The analysis is sufficiently different from that of Q-balls, primarily because typically only one defect per horizon is produced at the time of formation due to the Kibble mechanism~\cite{Kibble76}. However, the general mechanism remains the same: small number densities of defects lead to large fluctuations relative to the background density, these fluctuations become gravitationally bound and collapse to form black holes once the relic density has come to dominate, and the relics decay due to some instability (such as gravitational waves or decay to Nambu-Goldstone bosons in the case of cosmic strings). In order to accurately model production of PBHs from these defects, one should calculate the expected density perturbations on initially superhorizon scales, which only begin to grow once these scales pass back within the horizon and the defects come to dominate the universe's energy density.
Cosmic strings are probably the most likely candidate for primordial relics due to the fact that they are typically cosmologically safe, as the energy density in string loops is diluted during expansion at the same rate as radiation, $a^{-4}$ \cite{Vachaspati84,Turok1983}. 

We have shown that number density fluctuations of nontopological solitons in the early universe can be responsible for production of primordial black holes, and furthermore, that these black holes can make up all or part of the dark matter. Scalar fields and Q-ball formation are generic features of  supersymmetric extensions of the Standard Model, which provides a good motivation for this mechanism. Other scalar fields may exist and may undergo fragmentation, leading to PBH formation.  In addition, we have elucidated a possible mechanism through which topological defects may be able to produce primordial black holes as well under certain circumstances.

\textit{Acknowledgements.} This work was supported by the U.S. Department of Energy Grant No. DE-SC0009937.  A.K. was also supported by the World Premier International Research Center Initiative (WPI), MEXT, Japan.

\bibliography{pbh}
\end{document}